\documentclass[twoside,12pt]{article}
\usepackage{indentfirst}
\usepackage{bm}
\usepackage{graphicx}
\usepackage{amsmath}

\begin{document}    
\bibliographystyle{unsrt}

\begin{center}
\LARGE\bf Individuals' Mobility May Promote Criticality in Animal Collective Decision-making
\end{center}

\begin{center}
\rm Feng Hu*, Zhi Ting Wang, Fang Yang
\end{center}
\begin{center}
\begin{footnotesize} \sl
${\rm }$ \\  College of
Physics and Electronic Engineering, Chongqing Normal University, China. 

\end{footnotesize}
\end{center}

\vspace*{2mm}

\begin{center}
\begin{minipage}{15.5cm}
\parindent 20pt\footnotesize

It is highly believed that the individuals' mobility plays an
important role in phase transition in animal collective motion.
Here, we propose a model to study the effects of individuals'
mobility in a distributed animal collective decision-making process, during which each individual faces two options with equal quality.
We implement the quorum response rule, a type of social interaction rule which is
taxonomically recognized in animal collective decision-making, as
the sole interaction rule.
After the introduction of individuals' mobility, we find that the group can reach a consensus decision at one of the options at some critical points even the interaction is local. This result is an obvious contrast to the stationary individuals, the population of which is always equally distributed between the two options with fluctuations.  In order to explore the information dynamics, we introduce an important
information-theoretic measure, mutual information, to study the
critical behaviors. Furthermore, we study the case when individuals interact globally, and also find some qualitative similar critical behaviors.
\end{minipage}
\end{center}

\begin{center}
\begin{minipage}{15.5cm}
\begin{minipage}[t]{2.3cm}{\bf Keywords:}\end{minipage}
\begin{minipage}[t]{13.1cm}
collective animal behaviors; Self-organized critical behaviors; information
transfer;
\end{minipage}\par\vglue8pt
{\bf PACS: } 87.23.Cc, 89.70.Hj, 89.75.Da
\end{minipage}
\end{center}

\section{Introduction}
Collective animal behavior has captured many interests recently
because it exhibits complex cooperative behaviors
\cite{Sumpter2010collective}. It is now highly believed that
individuals' mobility plays an important role in the dynamic phase
transition from a disordered state to an
ordered one in animal collective motion\cite{Vicsek1995SPP}. Animal
collective decision-making, such as a colony of bees opting for a
nest \cite{Seeley2004bees}, an aggregation of cockroaches
choosing a shelter site \cite{Marc2006cockroach}, a school of fish
selecting a traveling path \cite{Ward2008fishes} \textit {etc.}, are examples that show some universal behaviors.
It is found that
information transfer and storage play an important role in animal
swarms' motion \cite{Wang2012InformationCascades, Herbert2015Initiation}.
Due to limited calculation capacity of
individual members, collective choice often results from large amount of local
interactions among individuals \cite{Couzin2007Collective}.
Interaction here means communication and action, which is the way by
which information propagates throughout the group.

Phase transition and critical behaviors are concepts well studied in
statistical physics \cite{Pathria2012Statistical} and are now widely
applied in dynamics of animal group behaviors \cite{Cavagna2010Scale,Buhl2006From,Beekman2001Phase}. It is now believed that
many biological systems, ranged from protein molecules in a cell to
organisms in a group, are poised at a critical
state\cite{Mora2011Are}. Because in critical state, variation of one
individual member correlates with that of any other ones in the
group no matter what the distance between them is. Thus the
correlation provides each animal with a much larger effective
perception range. Although mobility is a distinctive feature of animals, but
the important relationship between it and the group's decision-making capacity
is less studied.

Quorum response has been found working in many animal groups in the distributed collective decision process \cite{David2009quorumgeneral} including ant colonies
\cite{Mallon2001Individual}, cockroach aggregations
\cite{Am2006Collegial}, fish schools \cite{Ward2008Quorum} and
microbes \cite{Rossgillespie2014Collective}. It can also be applied in animal swarm motion and results in a dynamic phase transition from a disorder state to an ordered state \cite{Hu2016Application}.
It is essentially a positive feedback response, which can be exhibited in different math forms. Interestingly, it is observed that
ants can tune the parameters in quorum response rules to achieve a consensus
decision-making in some urgent situations, and the cost is that it may choose
an inferior option \cite{Pratt2006A, Chittka2009Speed}. The balance is a trad-off between efficiency and accuracy
that the group should be balanced.

In this paper, we propose a model to study the distributed animal collective
decision-making, and pay particular attention to the role played by individuates' mobility.
We implement the quorum response as the sole interaction rules. In the model, each individual selects locally from two options with equal quality. We first analyse the case when individuals are stationary and find that during the evolution the system, particles are equally distributed between two options with some fluctuations. We then introduce individuals' mobility into the group, and find that the group can make a consensus decision at one option as the nonlinearity of the quorum rules increase. In
order to explore the information dynamics, we
introduce mutual information to study the critical behavior.
Furthermore we compare the case when individuals move randomly with the
one when individuals interact globally. In both cases, information transfer rate is accelerated. And we find some qualitatively similar behaviors.

\section{An individual-based Model}
In the model, individuals of the group are categorized into three subgroups
of $x$, $y$ and $0$. Na\"{i}ve individuals in subgroup $0$ are searching for chances to become a member of subgroup $x$ or $y$. Once na\"{i}ve individuals  turn into subgroup $x$ or $y$, they will not turn back
to na\"{i}ve state anymore. This assumption is adapted and
simplified from a theoretical animal collective study \cite{David2009quorumgeneral}. Population
in subgroups $x$ and $y$ can also be transmitted from each other,
which is based on local quorum rules.
Eqn. (\ref{ModelEqn0}) to (\ref{ModelEqn2}) show the transitional probability  of individual $i$ depending on its local neighbors distribution in each subgroup at one time step.

\begin{eqnarray}
&& \label{ModelEqn0} dp_{i, 0} = - p \cdot p_{i, 0}\frac{n_{i,
x}^{k}}{n_{i, x}^{k}+T_{i}^{k}}- p \cdot p_{i,
0}\frac{n_{i, y}^{k}}{p_{i, y}^{k}+T_{i}^{k}},\\ && \label{ModelEqn1} dp_{i, x} = p \cdot p_{i, 0}\frac{n_{i,
x}^{k}}{n_{i, x}^{k}+T_{i}^{k}}+p_{i, y}\frac{n_{i, x}^{k}}{n_{i,
x}^{k}+T_{i}^{k}}-p_{i, x}\frac{n_{i, y}^{k}}{n_{i,
y}^{k}+T_{i}^{k}},\\ && \label{ModelEqn2}dp_{i, y} = p \cdot p_{i,
0}\frac{n_{i, y}^{k}}{p_{i, y}^{k}+T_{i}^{k}}+p_{i, x}\frac{n_{i,
y}^{k}}{n_{i, y}^{k}+T_{i}^{k}}-p_{i, y}\frac{n_{i, x}^{k}}{n_{i,
x}^{k}+T_{i}^{k}},
\end{eqnarray}
where $n_{i, 0}$, $n_{i, x}$ and $n_{i, y}$ are the population of $i's$ local neighbors in
subgroup $0$, $x$ and $y$, respectively. And $p_{i, 0}$, $p_{i, x}$ and $p_{i, y}$ are the probability of finding $i$ in  the corresponding subgroup $0$, $x$ and $y$, respectively. $p$ is a constant factor to limit na\"{i}ve individuals in subgroup $0$ to turn into subgroup $x$ or $y$ (we set $p\equiv\frac{1}{N}$, with $N$ being the whole population of the group). $T_{i}$ and $k$ are parameters in the quorum response
function (see eqn. (\ref{QREqn})). Eqn. (\ref{ModelEqn0}) shows the probability of individual $i$ transfer out of the subgroup $0$ to subgroup $x$ or $y$.  The first terms in Eqn. (\ref{ModelEqn1}) (or (\ref{ModelEqn2})) show the probability of $i$ transfer from subgroup $0$ into subgroup $x$ (or $y$). And the second and third terms show the probability of $i$ transfer between subgroup $x$ and $y$ in one time step. Note that $dp_{i, 0}+dp_{i, x}+dp_{i, y}=0$, which is required by the normalization of probability.

Here we adapt a simple Hill function form in quorum rules and implement it in the individual-based
model \cite{David2009quorumgeneral}, Eqn. (\ref{QREqn}) shows the propensity of an individual to join the subgroup $x$ ($y$).

\begin{equation}
\text{H}_{x,(y)}=\frac{n_{x,(y)}^{k}}{n_{x,(y)}^{k}+T^{k}}=\frac{(\frac{n_{x,(y)}/n}{T/n})^{k}}{(\frac{n_{x,(y)}/n}{T/n})^{k}+1},  \label{QREqn}
\end{equation}

\begin{figure}[h]
\includegraphics[width=12cm]{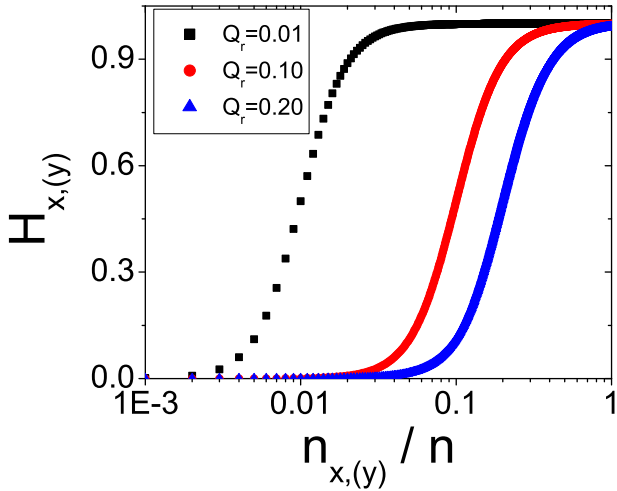}
\vspace{-0.2cm} \caption{(color online) Hill Function as quorum response (see Eqn.
(\ref{QREqn})). The black squares, red dots and blue triangles
represent the $Q_{r}$ of 0.01, 0.10 and 0.20,
respectively. We set $k=3$ in the figure. And the whole population of one individual's neighbors is fixed as $n=1000$. } \label{QuorumResponseFig}
\end{figure}

Fig. \ref{QuorumResponseFig} shows the character of the Hill function when $k=3$. We choose $T$ to be $10, 100, 200$ and $n$ to be $1000$ ($n$ could be seen as the whole population of neighbors). The crucial quantity
$Q_{r}\equiv \frac{T}{n}$ is defined as \textbf{quorum ratio}, since it correlates with an inflection point with a rapid
increase. Quorum response is an increase function and is sigmoid. The figure shows that when
the proportion of neighbors committed to option $x$ (or $y$) is less than the
quorum ratio, this individual will be less likely to choose this
option. On the contrary, if the proportion is larger than the quorum ratio, it
will be much more likely to choose this option.

Eqn. (\ref{ModelEqn0}) to (\ref{ModelEqn2}) are based on the individual-based quorum rules and capture the
mechanism of a positive feedback cycle that is often observed in
animal collective decision-making \cite{David2009quorumgeneral}. Note that there is also a relatively weak linear negative feedback cycle in the
equations, because the number of individuals who transfer out of a
subgroup is proportional to the population already committed to it.

In simulations of the individual-based model, initially we distributed total
$N$ ($\equiv1000$) individuals uniformly randomly on a line-world of length $L$ ($L\equiv20$).
And individuals commit to option $x$, $y$ and $0$ initially with
probability $I_{x}$, $I_{y}$ and $I_{0} $($I_{x}=I_{y}=0.1, I_{0}=0.8$), respectively. At each time
step, each individual searches the number of its neighbors within the search
radius $l$, then selects a subgroup based on quorum rules. The periodic boundary condition is applied. Each trial is simulated for a long time series ( 10000 steps), so eventually all the individuals are distributed at option $x$ or $y$ ($n_{x}+n_{y}=N$). We run 8 trials for each case.

\section{Results}

\begin{figure}[h]
\includegraphics[width=12cm]{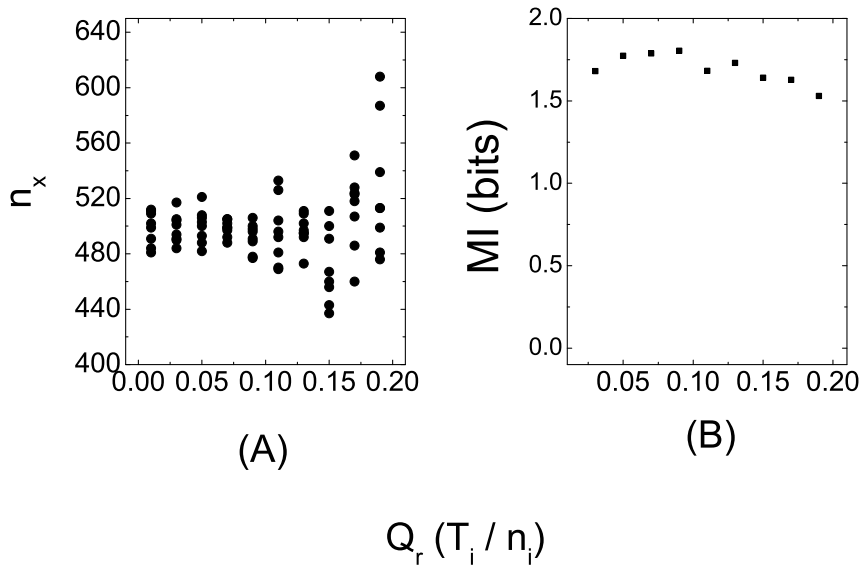}
\vspace{-0.2cm} \caption{Population dynamics (A) and average conditional mutual
information (B) versus quorum ratio. The individuals are stationary who
interact locally. The search radius of each individual is
$l= 0.1$.}\label{LocalStationary}
\end{figure}

Fig. \ref{LocalStationary} shows the group population dynamics versus the quorum ratio of
the stationary individuals who interact locally. In Fig. \ref{LocalStationary} (A),
the $y$-axis is the population in the whole group
who committed to option $x$ at last time step.
The data from different trials fluctuate around
the mean value of $N/2$ and don't settle in a stable value. We
believe because the interaction are local ($l=0.1$) and
the individuals are stationary, information flow is damped to local areas and it is impossible for the whole group
to reach a consensus decision.

We apply an important
information-theoretic measure, mutual information \cite{Cover2006Elements}, which measures the deviation from dependent distribution, to study the
information dynamics in the process. Eqn. (\ref{IndividualMI}) is individual $i's$  conditional mutual information, where $p(n_{i,x},n_{i,y},n_{i,0})$ is the joint probability whose neighbors in subgroup $x$, $y$ and $0$ are $n_{i,x}$,  $n_{i,y}$ and $n_{i,0}$, respectively. $p(n_{i,x},n_{i,y}|n_{i,0})$ is the conditional probability, whose subgroups have population of $n_{i,x}$ and $n_{i,y}$, given that the size of subgroup $0$ is $n_{i,0}$. The probability terms are calculated in the long time series from start to the last time step. Summation goes all the possible configurations of distribution of $n_{i}$, thus we let $n_{i,x}$, $n_{i,y}$ and $n_{i,0}$ runs from $0$ to $n_{i}$ separately. Then we average it over the whole population and trials to obtain the average conditional mutual information $MI$ (see Eqn. (\ref{AverageMI})).

\begin{eqnarray}
 && \label{IndividualMI} MI_{i}=\sum_{\substack{{n_{i,x}=0}\\n_{i,y}=0\\n_{i,0}=0}}^{n_{i}}
 p(n_{i,x},n_{i,y},n_{i,0}) \text {log}
 \frac{p(n_{i,x},n_{i,y}|n_{i,0})}{p(n_{i,x}|n_{i,0})p(n_{i,y}|n_{i,0})},
 n_{i}=n_{i,x}+n_{i,y}+n_{i,0}, \\
 && \label{AverageMI} MI=\sum_{trials}\sum_{i}(\frac{MI_{i}}{N})_{trials}
\end{eqnarray}

In Fig. \ref{LocalStationary} (B), the y-axis is $MI$, which captures the correlation between population size of
subgroup of $n_{x}$ and $n_{y}$ averagely, conditioned on the population
 size of subgroup $n_{0}$.  We see that the data are in a plateau, and there are no distinct features observed.

Next we introduce individuals' mobility to
the model. At each time step, each individual steps a distance
uniformly randomly chosen from $(-L, L)$ in the line-world.

\begin{figure}[h]
\includegraphics[width=12cm]{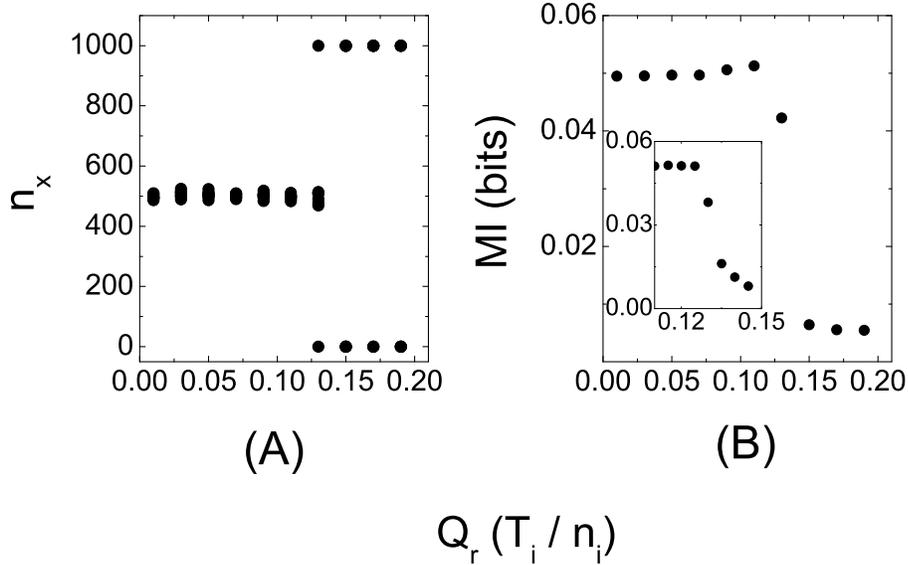}
\vspace{-0.2cm} \caption{Population dynamics (A) and average conditional mutual
information (B) versus  quorum ratio. The individuals interact
locally and the search radius of each individual is
$l= 0.1$, the same as in Fig.
\ref{LocalStationary}, except that each
individual steps a distance randomly uniformly chosen from $(-L,
L)$ at each time step. }\label{LocalMobility}
\end{figure}

Fig. \ref{LocalMobility} (A) shows the critical behaviors of the
population dynamics. Below the critical point around quorum ratio $0.12$,
the data from different trials show that populations select options $x$ (or $y$) with equal probability, so
the mean value is $N/2=500$. Beyond the critical point, symmetry is
broken and the whole population may reach a consensus decision at one option. The critical point corresponds to the initial population distribution at subgroups $x$, $y$, and $0$. When $Q_{r}<I_{x}$ (= $I_{y}=0.10$), quorum rules (see Eqn.\ref{QREqn}) show that the propensity for individuals transfer to subgroup $x$ or $y$ are both near to one. Thus the negative linear feedback in the model (see Eqn. (\ref{ModelEqn0}) to (\ref{ModelEqn2})) may result in the equal population distribution between the two options since they have same quality. As the quorum ratio increases and exceeds a critical value (i.e. $Q_{r}>I_{x}$), the positive feedback will take control the system evolution. The small fluctuations in the population distribution may be rapidly enhanced and result in the bifurcation of the system. Thus the whole group reaches a consensus decision at one of the options.

In Fig. \ref{LocalMobility} (B), we apply conditional mutual
information to measure the correlation between the subgroups of $x$
and $y$ conditioned on subgroup $0$, as we did in Fig. \ref{LocalStationary} (B).
We see that around the same critical point shown in Fig.
\ref{LocalMobility} (A), the mutual information drops to zero
rapidly. This shows that once the system begins phase transition, it proceeds in a very robust way and the population of one subgroup $x$ or $y$ has no correlation on the other. The inset shows a zoom in view to exhibit the critical
behaviors around the critical point.

Compared with the result in Fig. \ref{LocalStationary}, we believe the observed criticality results from the enhanced
information transfer due to individuals' mobility. Following this
line of thinking, if we let every individual interact globally,
which means information communication of individuals across the
group become instantaneous, the critical behaviors may reappear. So we run
the simulations with individuals interacting globally.

\begin{figure}[h]
\includegraphics[width=12cm]{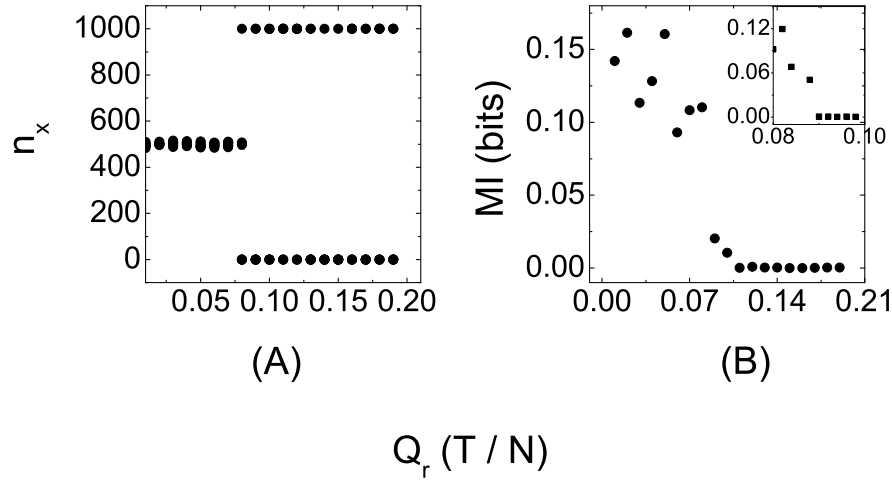}
\vspace{-0.2cm} \caption{Population dynamics (A) and conditional mutual
information (B) versus quorum ratio with global interacting individuals. } \label{GlobalFig}
\end{figure}

Fig. \ref{GlobalFig} shows the population dynamics and the conditional mutual
information behavior of global interacting individuals ($l=L$). In the figure the critical
point is around $0.1$, which is equal to the initial probability
distribution of $I_{x}=I_{y}=0.1$. When the threshold ratio is small
(i.e. $Q_{r} <0.1$), the negative feedback in Eqn.
(\ref{ModelEqn1}) and (\ref{ModelEqn2}) result in an equal
population-distribution over two options. When the quorum ratio grows bigger, the positive feedback takes a major control of the flow to subgroup of $x$ or $y$. We also see that in Fig. \ref{GlobalFig}(B), the plateau value of mutual
information is much higher than in Fig. \ref{LocalMobility} (B). Because in
global interacting case, when the quorum ratio is small, the
whole group rapidly distributes equally over two options.
Compared with the slow converging process of local interaction,
it makes the plateau value of mutual information
smaller.


\section{Summary and outlook}

We apply quorum response as interactions in our model to analyse the distributed animal collective
decision-making, given that each individual may choose from two options with equal quality. Quorum response interactions make the model showing features of strong positive feedback, as the quorum ratio becomes larger. In our model, we pay particular attention to the role of individuals' mobility and study two local interaction cases: stationary particles and randomly moving particles case. We find that individuals' mobility may pose the system to a critical state, in which the whole group may reach a consensus decision at one option if the quorum ratio exceeds a critical value. It is obviously in contrary to the stationary particle case, which shows no distinct features in evolution of the system as the quorum ratio varies. An important theoretic-information measure, mutual information, is applied to study the the correlation between the size of two subgroups distributed at the two options, conditioned on the size of the subgroup in which particles make no choice. We find that the mutual information drops rapidly from a plateau to ground, the reason of which, we believe, is resulted from the strong positive feedback features of the quorum response interactions and the enhanced information flow made by the moving particles. And we compare this case with the global interacting particles, in which information transfer is instantaneously, and find some qualitatively same critical behaviors.

An experiment exhibiting this kind of critical behavior was observed when
cockroach aggregations selecting from two shelter sites with equal quality
\cite{Am2006Collegial}. It was found that as the shelters' carrying
capacity increase to a critical value, the equally distributed
population over two sites suddenly shifts to one of them. In order
to explain the experiment, it was suggested in the paper that each individual could perceive the whole population distribution in the experiment setup.
In our paper, we find that individuals' mobility can enhance the rate of information flow of the local interacting particles and poise the system to a critical state, as the global interacting individuals do. Another human experiment found that redundant connections could increase the behavior spread on line, because one person's behavior variation needs certain amount influences from peers \cite{Olver2010Spread}. In our work, we find that individuals' mobility plays an important role into making the whole group to reach a consensus decision given that the quorum ratio reaches a limit.

Collective animal groups, such as starling flocks, fish schools or
ant colonies \textit {etc.}, are recently being called ``collective
minds''\cite{Couzin2007Collective}, because of
the efficient information processing ability. We have introduced mutual information to study the
the critical dynamics in this paper and more information-theoretic
measures are expected \cite{Schreiber2000Measuring} to explore the distributed animal collective decision-making behaviors in future.

\section*{Acknowledgement}

Project supported by the Science and Technology
Research Program of Chongqing Municipal Education Commission (Grant No.
KJ1703061)

\vspace*{4mm}
* fenghu@cqnu.edu.cn

\bibliography{bibfiles}

\end{document}